\documentclass[fleqn,usenatbib]{mnras}
\usepackage{newtxtext,newtxmath}
\usepackage[T1]{fontenc}

\DeclareRobustCommand{\VAN}[3]{#2}
\let\VANthebibliography\thebibliography
\def\thebibliography{\DeclareRobustCommand{\VAN}[3]{##3}\VANthebibliography}

\usepackage{setspace}

\usepackage{graphicx}	
\usepackage{amsmath}	

\usepackage{bm, graphicx, subfigure, amsmath, morefloats, mathtools, amsfonts}
\usepackage{color}
\hypersetup{allcolors = [rgb]{0., 0., 0.5}}

\usepackage{array, booktabs, makecell, multirow}  

\usepackage{longtable}

\newcommand{\code}[1]{\texttt{#1}}
\newcommand{\documentname}

\newcommand{\teff}{\mbox{$T_{\rm eff}$}}

\newcommand{\feh}{\mbox{$\rm [Fe/H]$}}

\newcommand{\alphafe}{\mbox{$\rm [\alpha/Fe]$}}

\newcommand{\logg}{\mbox{$\log g$}}

\newcommand{\Rbirth}{$R_{\rm birth}$}

\usepackage{xcolor}
\definecolor{pink}{rgb}{0.98, 0.38, 0.5}

\title[Abundance distribution across birth radii]{The individual abundance distributions of disc stars across birth radii in GALAH}

\author[K. Wang et al.]{
Kaile Wang,$^{1}$\thanks{E-mail: kailewang@utexas.edu}
Andreia Carrillo,$^{2}$
Melissa K. Ness,$^{3,4}$
Tobias Buck$^{5}$
\\
$^{1}$Department of Astronomy, University of Texas at Austin, 2515 Speedway, Stop C1400, Austin, TX 78712-1205, USA\\
$^{2}$Institute for Computational Cosmology, Department of Physics, Durham University, Durham DH1 3LE, UK\\
$^{3}$Department of Astronomy, Columbia University, Pupin Physics Laboratories, New York, NY 10027, USA\\
$^{4}$Center for Computational Astrophysics, Flatiron Institute, 162 5th Avenue, New York, NY 10010, USA\\
$^{5}$Leibniz-Institut f{\"u}r Astrophysik Potsdam (AIP), An der Sternwarte 16, D-14482 Potsdam, Germany
}

\date{Accepted XXX. Received YYY; in original form ZZZ}

\pubyear{2023}

\begin{document}
\label{firstpage}
\pagerange{\pageref{firstpage}--\pageref{lastpage}}
\maketitle

\begin{abstract}
Individual abundances in the Milky Way disc record stellar birth properties (e.g. age, birth radius (\Rbirth)) and capture the diversity of the star-forming environments over time. Assuming an analytical relationship between ([Fe/H], [$\alpha$/Fe]) and \Rbirth, we examine the distributions of individual abundances [X/Fe] of elements C, O, Mg, Si, Ca ($\alpha$), Al (odd-z), Mn (iron-peak), Y, and Ba (neutron-capture) for stars in the Milky Way. We want to understand how these elements might differentiate environments across the disc. We assign tracks of \Rbirth\ in the [$\alpha$/Fe] vs. [Fe/H] plane as informed by expectations from simulations for $\sim 59,000$ GALAH stars in the solar neighborhood ($R\sim7-9$ kpc) which also have inferred ages. Our formalism for \Rbirth\ shows that older stars ($\sim$10 Gyrs) have a \Rbirth\ distribution with smaller mean values (i.e., $\bar{R}_{\mbox{birth}}$$\sim5\pm0.8$ kpc) compared to younger stars ($\sim6$ Gyrs; $\bar{R}_{\mbox{birth}}$$\sim10\pm1.5$ kpc), for a given [Fe/H], consistent with inside-out growth. The $\alpha$-, odd-z, and iron-peak element abundances decrease as a function of \Rbirth, whereas the neutron-capture abundances increase. The \Rbirth-[Fe/H] gradient we measure is steeper compared to the present-day gradient (-0.067 dex/kpc vs -0.058 dex/kpc), which we also find true for \Rbirth-[X/Fe] gradients. These results (i) showcase the feasibility of relating the birth radius of stars to their element abundances,  (ii) the abundance gradients across \Rbirth\ are steeper than those over current radius, and (iii) offer an observational comparison to expectations on element abundance distributions from hydrodynamical simulations.
\end{abstract}


\section{Introduction}
Recovering the birth conditions of the stars is one of the main goals of Galactic archaeology. However, stars deviate from their birth orbits, such that their guiding-center radius can change over their lifetime, without leaving any signature of this change. These orbital excursions are due to processes such as the interaction with the spiral structure as well as external perturbations from infalling satellites (e.g. \citealt{Sellwood2002, MinchevQuillen2006, Quillen2009, Minchev2012}). Although we cannot directly probe the initial  orbital properties of disc stars at birth, they exhibit atmospheric abundances that - to first order -  reflect the abundance distribution of the gas from which the stars were born, with exceptions (e.g. \citealt{Iben1965, Bahcall1990, Dotter2017, Shetrone2019}).

We can therefore assume that most element abundances of stars, in particular within narrow regions of evolutionary state, are time-invariant. With stellar death, elements created within the stars and during explosive nucleosynthesis are returned to the interstellar medium. This  enriches the environment where newer stars are formed, in a cyclic process. The element abundances for a given star are therefore a record of the nucleosynthetic history of the star-forming environment, at that particular time and place. The time invariance of element abundances and their effective barcode of a star's birth environment has been foundational to the idea of chemical tagging, via which individual molecular cloud stellar birth sites in the disc might be reconstructed using abundances alone \citep{Freeman&BH2002}. However, the current data appear to demonstrate that this goal is prohibited by the low-dimensionality of what appears to be a very correlated abundance space
\citep{Ting2015, Kos2018, Ness2019, deMijolla2021, Griffith2021, Ness2022, Weinberg2022}. A more feasible goal with current spectroscopic data is the inference of the time and overall radius at which stars formed in the disc.

Different types of stars and production mechanisms produce elements across the periodic table with different yields, at different rates, and at different points in time (see \citealt{Kobayashi2020}). Additionally, it is widely accepted that galaxies, like the Milky Way, formed inside-out, with star formation starting in the deepest part of the potential and proceeding outwards (e.g. \citealt{GonzalezDelgado2015,Frankel2019}). Combining nucleosynthesis timescales with the inside-out formation of the Milky Way, the element abundances of the stars encode the temporal enrichment of the Galaxy and reveal stars' birth properties in terms of age and spacial location. We are now able to have a clearer picture of this as the field of Galactic archaeology has greatly expanded due to large multi-object stellar surveys, such as the Apache Point Observatory Galactic Evolution Experiment (APOGEE; \citealt{Majewski2017, Abdurro2022}), the Galactic Archaeology with HERMES (GALAH; \citealt{Buder2021}), Gaia-European Southern Observatory (ESO) survey (Gaia-ESO; \citealt{Gilmore2012}), and the Large Sky Area Multi-Object Fibre Spectroscopic Telescope (LAMOST; \citealt{cui12, Zhao2012}). These surveys enable the detailed study of the element abundance for $>10^{5}$ stars in the Galaxy. 

In addition to element abundances, another fundamental and time invariant property of stars is their age. Age tells us \textit{when} during the evolution of the galaxy a star was formed. In fact, numerous studies have explored the relationship between stellar age and element abundances, or age-[X/Fe] relations (\citealt{Hayden2020, Carrillo2022, RatcliffeNess2022}). These studies have shown that just by knowing a star's age and metallicity, [Fe/H], the element abundance, [X/Fe], can be predicted up to a precision of 0.02 dex for many elements. Indeed, a star's age does prove to be a key link to understanding the nucleosynthetic history and evolution of the Galaxy.
The only missing link now is \textit{where} in the galaxy a star was born. If we know a star's individual abundances, age, and birth site, we can begin to unravel the formation of the Milky Way disc with utmost detail.

To this effect, the element abundances of stars should be yet very useful. Earlier works have demonstrated the feasibility to infer birth radius, \Rbirth. For example, \citet{Minchev2018} presented a largely model-independent approach for estimating \Rbirth~for Milky Way disc stars, using [Fe/H] and age estimates from the local HARPS sample \citep{Adibekyan2012}. The assumptions relied on are (1) the interstellar medium (ISM) is well mixed at a given radius, (2) there exists a negative radial metallicity gradient in the ISM for most of the disc lifetime, (3) stars younger than 1~Gyr are expected to have little migration, and (4) the Milky Way formed inside-out. Utilizing the \Rbirth~derived in their work, they find that the ISM radial metallicity gradient in the Milky Way disc flattens with time. As noted in this study, processes like radial migration can blur \Rbirth~signatures. With this in mind, \citet{Frankel2018} developed a model to derive \Rbirth~by quantifying the radial migration in the Milky Way disc, using the ages and [Fe/H] of low-$\alpha$ disc stars. In this work, it was assumed that (i) the metallicity of the ISM has negligible variations azimuthally, (ii) the Milky Way had a relatively quiescent life for the past 8~Gyr, and (iii) radial orbit migration is the only mechanism responsible for the scatter in age–metallicity at a given radius. Their model reproduced the observed data well and further found that the radial orbit migration efficiency in the Milky Way is strong. Recently, \citet{Lu2023} proposed an empirical method to derive birth radii from age and metallicity measurements, with the assumptions that gas is well mixed in Galactic azimuth, Milky Way formed inside-out, and there is a well-defined linear relation between metallicity and birth radius. Such in-depth studies to derive \Rbirth~have been shown to be successful, with the help of various physically-motivated assumptions and modeling. It is therefore worth asking if \Rbirth~could be similarly derived with different assumptions, and specifically a model that does not directly use present-day radius measurements. In addition, as detailed element abundances have been shown to have a direct link to ages, it is interesting to explore how detailed element abundances can potentially trace stars back to their birth sites. 

Fortunately, the correlations between the birth radius and stellar properties have also been shown from cosmological hydrodynamical simulations, allowing methods of recovering the birth radius of stars to be explored. For example, \citet{Lu2022a} examined the reliability of inferring birth radii from the assumed linear relationship between the ISM metallicity with radius, using four zoom-in cosmological hydrodynamic simulations from the NIHAO-UHD project (\citealt{Buck2020,Buck2020b}). They found that precise stellar birth radii can be obtained for stars with age $< 10$ Gyr, as the stellar disc starts to form and the linear correlation between the ISM metallicity and radius increases. Also with the simulations from the NIHAO-UHD project, \citet{Buck2020} showed the direct correlation between element abundances (specifically, [O/Fe] and \feh) and the birth location of stars.

In this work, we want to recover the birth radius of stars simply based on their \alphafe \ and \feh \ abundances, as shown in simulation works (e.g. \citealt{Buck2020}). Instead of performing complex Galactic chemical evolution modeling, we assign each of the stars a birth radius based on their \alphafe \ and \feh \ abundances and examine the validity of this birth radius assignment. To do this, we explore the individual abundance distribution [X/Fe] across birth radii with the disc stars in GALAH DR3 data \citep{Buder2021}. In Section \ref{Data}, we describe the observational data we used in this study. In Section~\ref{method}, we discuss our birth radius assignment and the simulation work we are motivated by. In Section~\ref{AgeMetallicity}, we present our age-birth radius relation in two thin metallicity bins, and in Section~\ref{eledisdis} we show the distribution of individual element abundances [X/Fe] across birth radii. The results presented in these two sections validate our birth radius assignment based on element abundances. Lastly, we summarize and discuss the results in Section~\ref{Discussion}.

\section{Observational Data}
\label{Data}
We take advantage of the Galactic Archaeology with HERMES (GALAH) survey data release 3 (DR3, \citealt{Buder2021}) which measures up to 30 element abundance ratios for elements in different groups: $\alpha$, light/odd-z, iron-peak, and neutron-capture. The GALAH survey uses the HERMES instrument, a high-resolution (R~$\sim~28,000$) four channel fibre-fed spectrograph (covering 4713–4903 \AA, 5648–5873 \AA, 6478–6737 \AA, and 7585–7887 \AA) on the Anglo-Australian Telescope \citep{DeSilva2015}. The catalogue contains 588,571 stars, with the stellar parameters determined using the modified version of the spectrum synthesis code Spectroscopy Made Easy (SME: \citealt{VP1996}; \citealt{PV2017}) and 1D MARCS model atmospheres. After the stellar parameters were estimated and fixed, one abundance was fitted at a time for the different lines/elements in the GALAH wavelength range \citep{Buder2021}. In this work, we aim to study the distribution of individual abundances [X/Fe], which we take to be X = $\alpha$, C, O, Mg, Al, Si, Ca, Mn, Y, and Ba, spanning the different groups of elements.

In addition to the main catalogue, we also use the GALAH DR3 value-added catalogue that contains stellar ages, Galactic kinematics, and dynamics. The stellar ages were determined by the Bayesian Stellar Parameter Estimation code (BSTEP), an isochrone-based
scheme that provides a Bayesian estimate of intrinsic stellar parameters from observed parameters by making use of stellar isochrones, adopting a flat prior on age and metallicity \citep{Sharma2018}. The Galactic dynamic information was calculated using \code{galpy} \citep{Bovy2015}. In the calculations, the best fitting axisymmetric potential by \citet{McMillan2017} was used with a Solar radius of 8.21 kpc.

\begin{figure}
\centering
\includegraphics[width=0.48\textwidth]{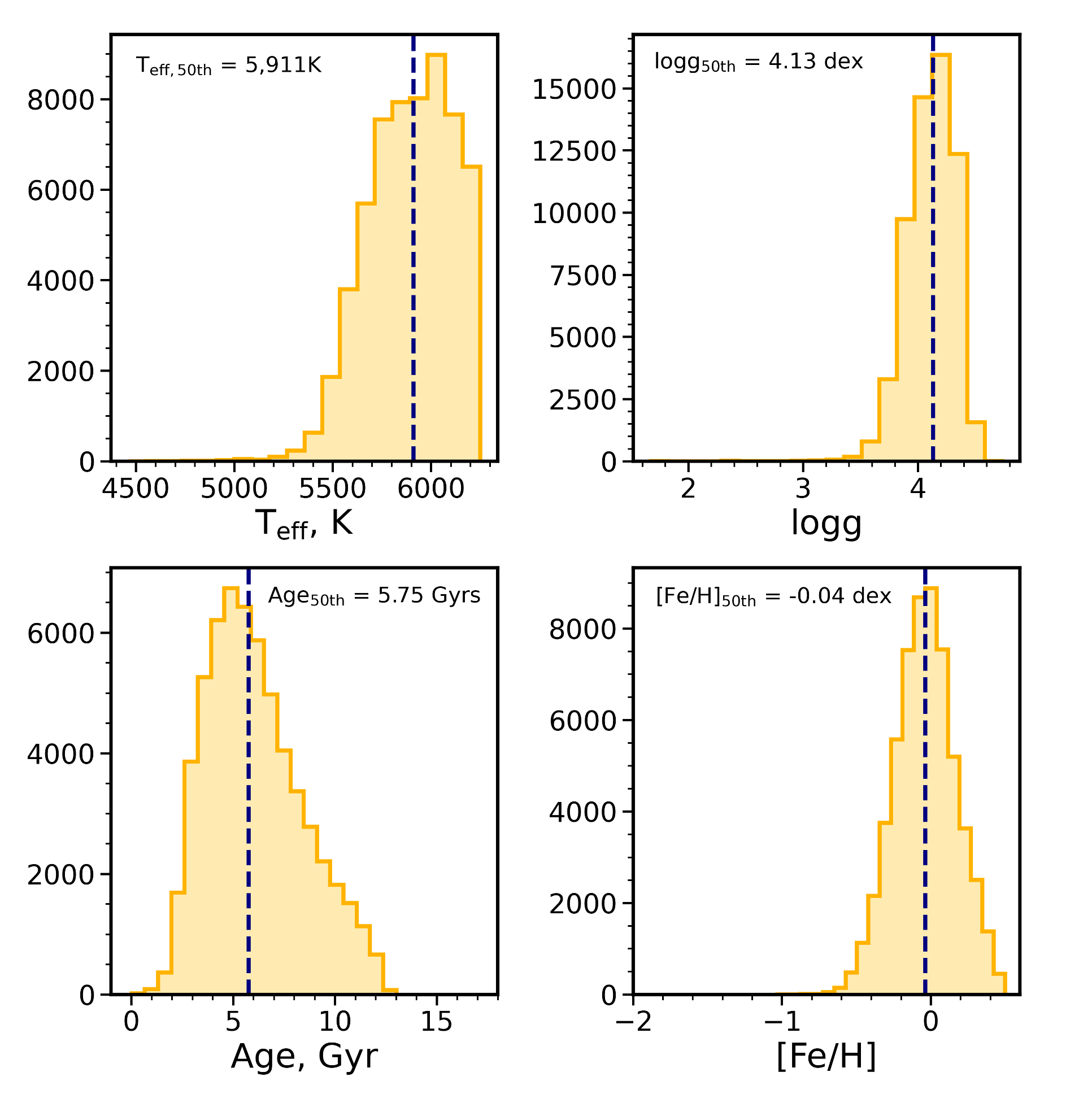}
\caption{Distribution of stellar parameters (effective temperature \teff, surface gravity \logg, stellar age, and metallicity \feh) for the parent sample of 59,124 qualified disc stars. The vertical dashed lines mark the median of the distributions.} 
\label{fig:StellarPara}
\end{figure}

\begin{figure}
\centering
\includegraphics[width=0.48\textwidth]{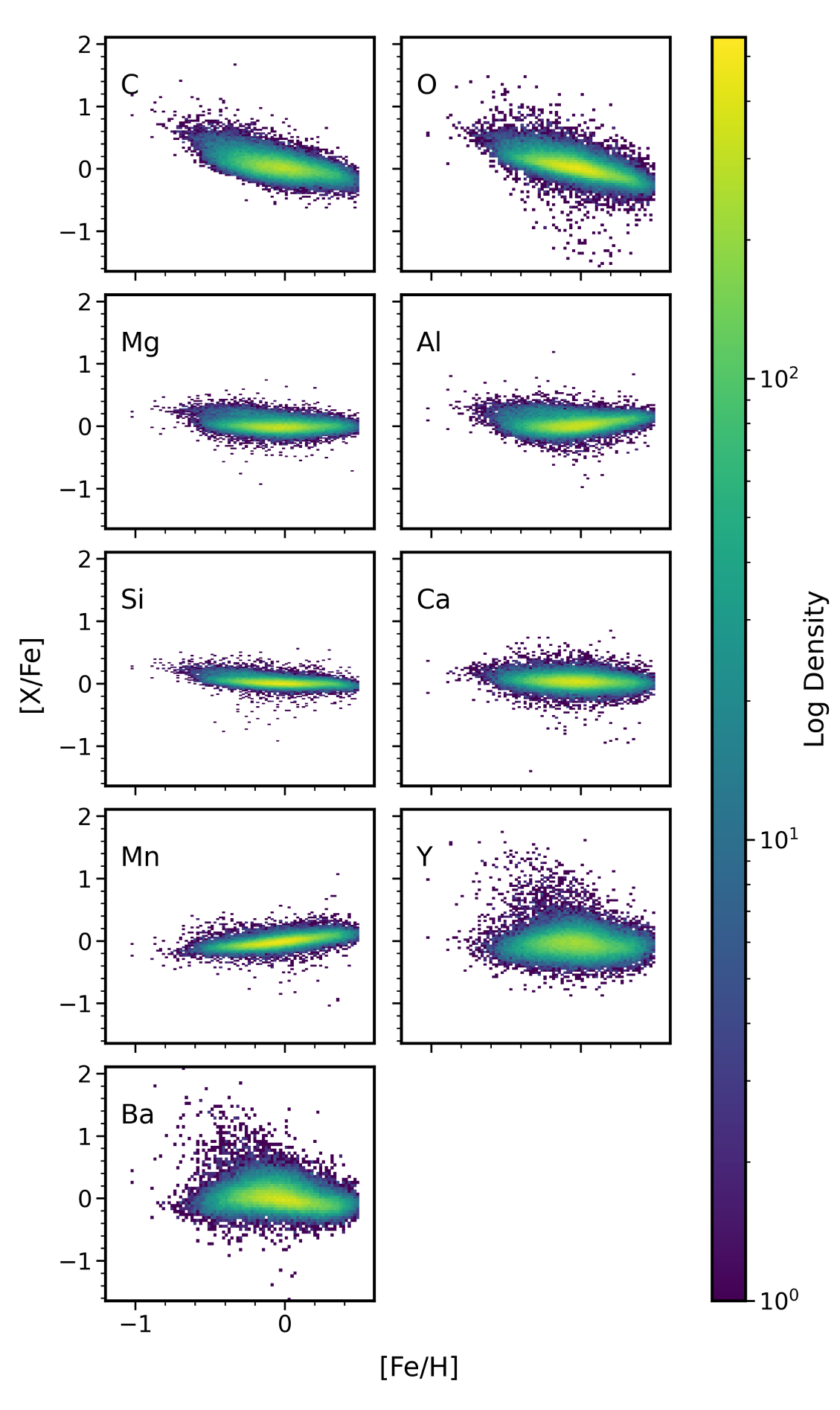}
\caption{C, O, Mg, Al, Si, Ca, Mn, Y, and Ba abundances of the qualified parent sample of 59,124 stars as a function of metallicity [Fe/H], colored by log density.} 
\label{fig:Xfe-Feh}
\end{figure}

We assemble a parent sample of qualified GALAH DR3 disc stars according to the following criteria:
\begin{center}
\begin{itemize}
\centering
\item  flag\_sp=0, flag\_fe\_h=0, flag\_X\_fe=0
\item  $-2 < \text{[Fe/H], \alphafe} < 0.5$, $-1 <$ \logg$ <6 $
\item  $3500 < \teff < 6250$ K, SNR = snr\_c3\_iraf $> 40$
\item  $7 < \text{R} < 9$, and $|\text{z}| < 2$
\end{itemize}
\end{center}

\noindent where X = $\alpha$, C, O, Mg, Al, Si, Ca, Mn, Y, and Ba. We set the cut in element abundance to avoid extreme values. The flag\_sp, flag\_fe\_h, and flag\_X\_fe are set to select stars with reliable stellar parameters and element abundance determination. In addition, we limit the \teff \ range such that the abundances are not affected by systematic temperature trends. This selection produces agreement between the \teff \ values from GALAH+ DR3 and from angular diameter-based measurements (e.g. \citealt{Karovicova2018, Karovicova2020}) for Gaia benchmark stars \citep{Buder2021}. We show in the Appendix in Figure \ref{fig:RunnningMean} that there are only small slopes between element abundances [X/Fe] and \teff. These may be real or systematics inherited from stellar models. We employ a signal-to-noise ratio cut of SNR $>$ 40 for the red band (CCD 3) to ensure good quality spectra, as well as cuts in Galactocentric radius (R) and height from the disc plane (z) to select for disc stars. This results in a sample of 59,124 stars, and the stellar parameters are shown in Figure~\ref{fig:StellarPara}. The stars span a range of 0.004 to 13 Gyrs in age, with a median age = 5.7 Gyrs. The 16th and 84th percentiles of age are 3.7 Gyrs and 8.6 Gyrs respectively. Figure \ref{fig:Xfe-Feh} shows the density plots of the parent sample on [Fe/H]-[X/Fe] plane, for elements C, O, Mg, Al, Si, Ca, Mn, Y, and Ba.

\section{Method}
\label{method}
\begin{figure*}
    \centering
    \includegraphics[width=0.8\textwidth]{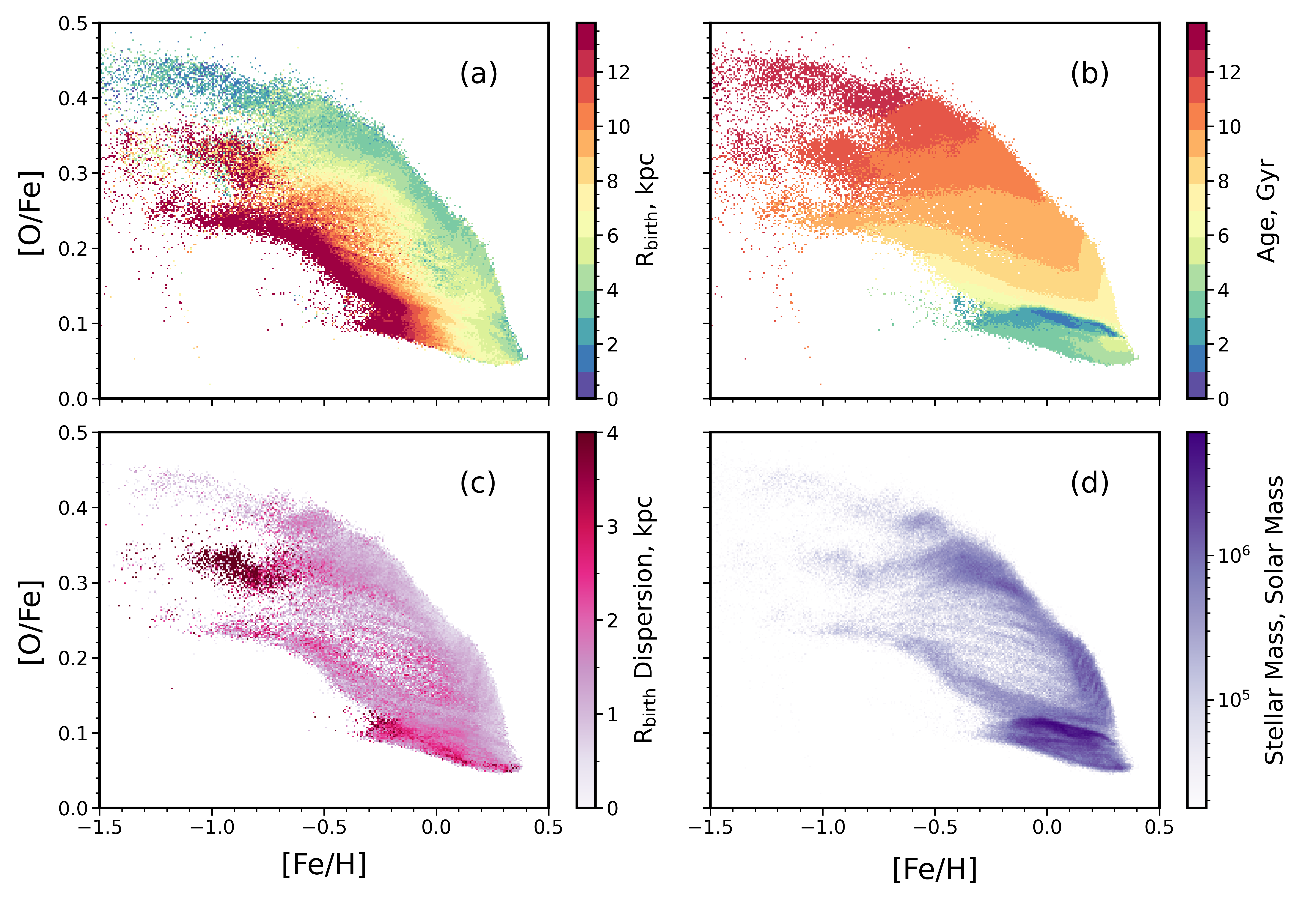}
    \caption{Reproduction of Figure 3 in Buck 2020 for the galaxy g2.79e12, showing the [O/Fe] vs. \feh \ plane at the solar radius (7 $<$ R $<$ 9 kpc) colored by (a) birth radius, (b) age, (c) birth radius dispersion, and (d) stellar mass.} 
    \label{fig:Buck2020}
\end{figure*}

\begin{figure*}
\centering
\includegraphics[width=\textwidth]{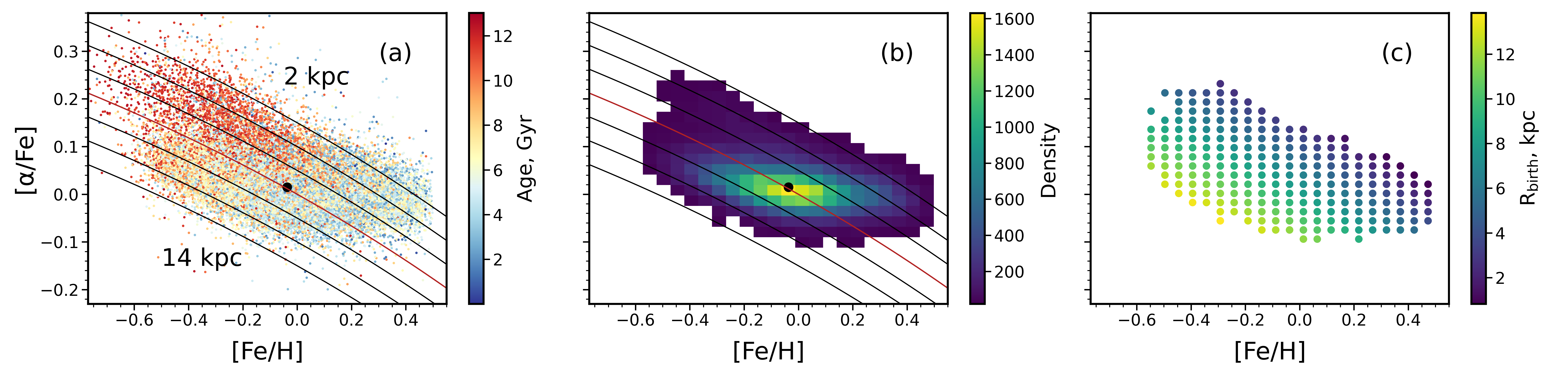}
\caption{(a) Distribution of the parent sample on the [$\alpha$/Fe] vs [Fe/H]  plane colored by age. The data covers a region of the stars between $7 < R < 9$ kpc and $|z| < 2$ kpc. The median [$\alpha$/Fe] and [Fe/H] of the population is indicated by the black point. The curves represent the birth radii assigned to the population. The birth radius increases as going from the top to the bottom curve (top curve: \Rbirth \ = 2 kpc; bottom curve: \Rbirth \ = 14 kpc). Specifically, the red curve indicates the median assigned birth radius (8 kpc). (b) Density plot of the parent sample. All bins that have a count less than 20 stars are removed. Birth radius tracks are laid down, and again the median ([$\alpha$/Fe], [Fe/H]) is shown by the black dot. (c) Binned data distribution on [$\alpha$/Fe] vs [Fe/H] plane, colored by mean \Rbirth. Stars with assigned \Rbirth \ $<$ 0 and bins with $<$ 20 stars are removed.}
\label{fig:GalahDist}
\end{figure*}

We aim to determine, given \alphafe \ and \feh \  abundances, the distribution of [X/Fe] across different birth radii (\Rbirth), under an assumed relation between [Fe/H]-[$\alpha$/Fe] and \Rbirth. Cosmological simulations (e.g. \citealt{Buck2020}) demonstrate clear birth radius tracks on the [O/Fe] vs. \feh \  abundance plane. Figure~\ref{fig:Buck2020} is a reproduction of Figure 3 in \citet{Buck2020} for the galaxy g2.79e12, showing the [O/Fe] vs. \feh \ plane at solar radius ($7<R<9$~kpc). The zoom-in simulation of g2.79e12 analyzed in \citet{Buck2020} is taken from the Numerical Investigation of a Hundred Astronomical Objects (NIHAO) simulation suite of cosmological hydrodynamical simulations of Milky Way mass galaxies (\citealt{Wang2015,Buck2020b}). The total virial mass, total stellar mass, and the disc scale length of g2.79e12 are $3.13\times{10^{12}}{M_{\odot}}$, $15.9\times{10^{10}}{M_{\odot}}$, and 5.57 kpc. Figure \ref{fig:Buck2020} panels are colored by (a) birth radius, (b) age, (c) birth radius dispersion, and (d) stellar mass. In Figure~\ref{fig:Buck2020} panel (a), stars with high [O/Fe] (>0.3) are seen to mostly originate from the inner Galaxy, while stars with low [O/Fe] (< 0.2) are distributed across a wider range of birth radii where larger birth radii are offset to lower metallicity. Panel (b) shows clear horizontal age gradients with older ages associated with higher [O/Fe]. In panel (c), there is high birth radius dispersion around \feh \ = -1.0, [O/Fe] = 0.3, as well as the lower-right region on the [O/Fe] vs. \feh \ plane towards high metallicity. The stellar mass is also higher in the lower-right region, as shown in panel~(d).
 
Motivated by the results from the \citet{Buck2020} simulations, specifically the birth radius-element abundances trends, we lay down seven \Rbirth\ tracks (2, 4, 6, 8, 10, 12, and 14 kpc) as shown in Figure \ref{fig:GalahDist} panel (a) in the \alphafe \ vs. \feh \ plane from GALAH data. These tracks can be described by the following equation
$$\text{\Rbirth} = -40\times(\text{\alphafe}+0.80\times \exp{(0.4\times\text{\feh})}-0.81)+8$$
which was obtained by fitting birth radius tracks similar to Figure \ref{fig:Buck2020} panel (a). We further assign every star in our sample a birth radius according to the equation, with known \feh\ and \alphafe. The number of stars in bins between each track is as follows: 3804 (2-4 kpc), 9328 (4-6 kpc), 15720 (6-8 kpc), 18628 (8-10 kpc), 9565 (10-12 kpc), 1306 (12-14 kpc). Stars with assigned \Rbirth \ $<$ 0 kpc are removed (170 stars, or 0.29\% of all qualified disc stars).

Instead of using the oxygen abundance [O/Fe], we choose to use the alpha element abundance because (1) it is better measured, as the mean uncertainty in \alphafe \ is smaller than that of [O/Fe], and (2) in the simulations performed by \citet{Buck2020}, [O/Fe] is intended as a tracer of alpha-elements than of the specific element O. The absolute values of each radial track are not calibrated to match the Milky Way, but the range is consistent with the birth radius range used in other studies, e.g. \citealt{Frankel2018}. We adopt this form of the relation between [Fe/H]-[$\alpha$/Fe]-birth radius and examine the overall effect for the birth radius variations in the element abundance distributions, and for the birth radius at fixed age, if such a relation is held in the Milky Way. 

The birth radius increases as [$\alpha$/Fe] decreases (from top to bottom), and the y-axis spacing between two neighboring \Rbirth\ curves is around 0.05 dex. The distribution of the parent sample stars on the [$\alpha$/Fe] vs. [Fe/H] are also shown in Figure \ref{fig:GalahDist} panel (a) colored by age, with the median (\feh,\ \alphafe) shown as a black circle. We lay down the \Rbirth\ tracks such that the middle \Rbirth\ track goes over the median (\feh,\ \alphafe) point because most of the stars are located near the origin, i.e. ([Fe/H], [$\alpha$/Fe]) = (0, 0) (see density plot in Figure \ref{fig:GalahDist} panel (b)). Furthermore, the birth radius for the majority of stars roughly follows a similar distribution as their current Galactocentric radii \citep{Carrillo2022}, which is around 8 kpc. Additionally, the distribution of stellar ages exhibits a decreasing trend going towards lower \alphafe\ and higher \feh as shown in Figure \ref{fig:GalahDist} panel (a).

 As shown in Figure \ref{fig:GalahDist} panel b, the stellar population density is very non-uniformly distributed in the \feh-\alphafe\ plane. We wish to carry out an analysis of how the age and individual abundance distributions of stars change with  birth radius, given our \Rbirth\ model assigned in the \feh-\alphafe\ plane. Therefore, the varying density distribution of stars in this plane is not information we wish to propagate. To eliminate the impact of the uneven density of stars in the \feh-\alphafe\ plane for this analysis, we use a grid of evenly spaced representative populations in [$\alpha$/Fe] vs. [Fe/H]. Along the x and y axis, the grid spacing is 0.051 and 0.019, respectively. Bins with N~$<$~20 stars are removed, mostly on the edges, because we want our binned data to be representative of the neighboring star population on the abundance plane.

The remaining sample of 231 binned data points, including 57,858 stars, is summarised in Figure \ref{fig:GalahDist} panel (c) colored by mean birth radius. We use these binned data points, which give us an even sampling across the \feh-\alphafe plane in \Rbirth, for further analysis.
 
\section{Birth Radius distributions with age and metallicity}
\label{AgeMetallicity}
\begin{figure*}
\centering
\includegraphics[width=\textwidth]{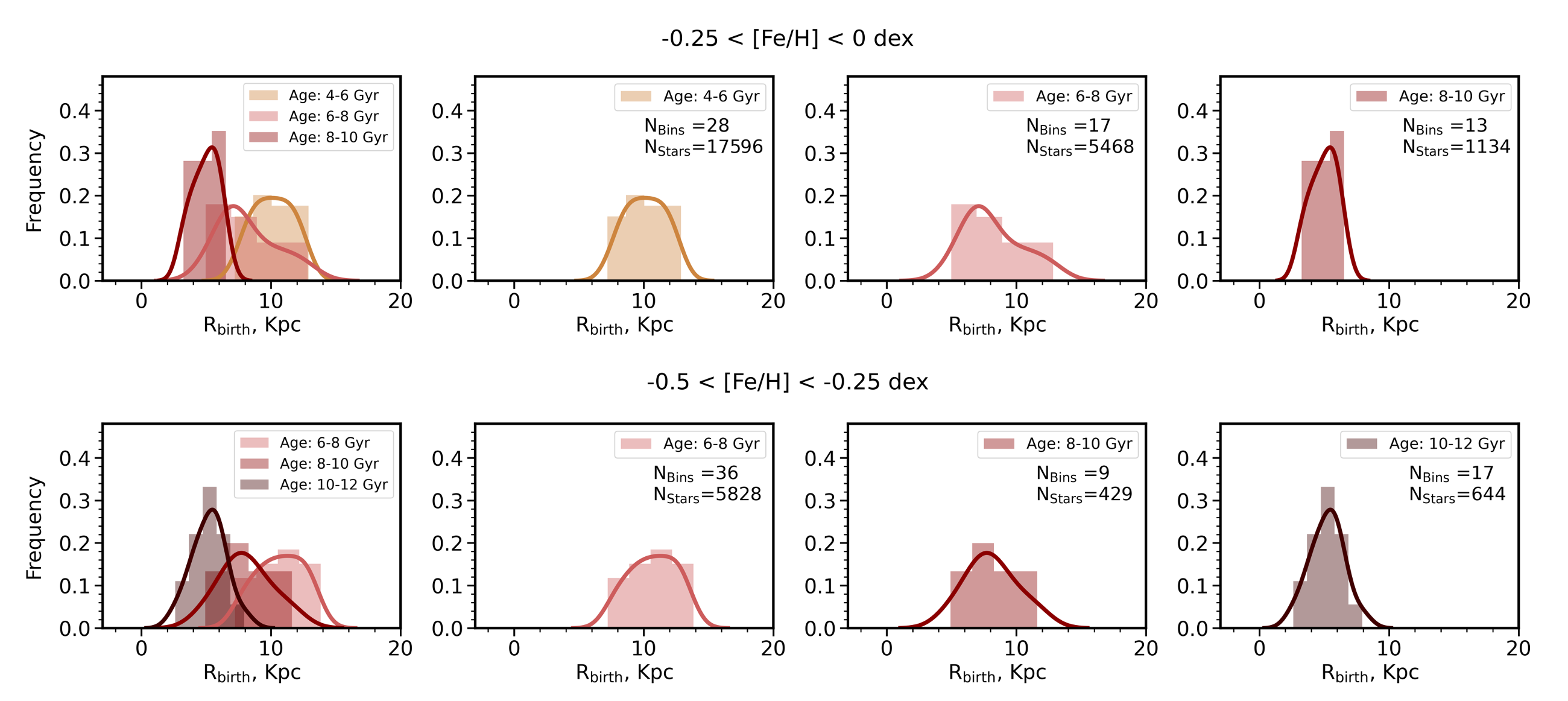} 
\caption{\Rbirth\ kernel density estimations (KDEs) of high (-0.25 $<$[Fe/H]$<$0, top panel) and low (-0.5$<$[Fe/H]$<$-0.25, bottom panel) metallicity stars as sampled from Figure 4(c). The location of the peaks shows that older stars have a smaller mean birth radius than younger stars, indicating the inside-out formation of the Milky Way. The leftmost sub-panels are showing the \Rbirth\ distributions of the three age bins at the same time, whereas the other six sub-panels are individually showing the \Rbirth\ distributions of different age bins.} 
\label{fig:MetDist}
\end{figure*}

We explore how the birth radius distribution of stars in the \feh-\alphafe\ plane as shown in Figure 4 (c) changes as a function of age and metallicity. In Figure \ref{fig:MetDist}, we show the birth radius distribution for a high metallicity (-0.25$<$[Fe/H]$<$0, top panel) and low metallicity (-0.5$<$[Fe/H]$<$-0.25, bottom panel) sample. Within the same metallicity bin, the sample is broken down into three stellar ages bins. These are shown separately with different colors in the sub-panels of Figure \ref{fig:MetDist}, with the lightest to darkest color for the youngest to oldest stars, respectively.

The mean birth radius values for the three age bins lie at 10.1 kpc (4-6 Gyr bin), 8.2 kpc (6-8 Gyr bin), and 5.0 kpc (8-10 Gyr bin) for the high metallicity sample, and at 10.7 kpc (6-8 Gyr bin), 8.1 kpc (8-10 Gyr bin), and 5.2 kpc (10-12 Gyr bin) for the low metallicity sample. For both the high and low metallicity samples, the birth radius distribution for older stars generally peaks at a smaller birth radius compared to younger stars, exhibiting an inside-out formation trend similar to other studies (e.g. \citealt{Minchev2018,Queiroz2020,Carrillo2020}). Furthermore, the width of the birth radius distributions also has a correlation with age, in which the width decreases with increasing age. The median absolute deviations (MAD) of the three high metallicity age bins are 1.4 kpc (4-6 Gyr bin), 1.2 kpc (6-8 Gyr bin), and 0.8 kpc (8-10 Gyr bin), and the values for the low metallicity sample are 1.5 kpc (6-8 Gyr bin), 1.2 kpc (8-10 Gyr bin), and 0.8 kpc (10-12 Gyr bin). Here we choose MAD to describe the dispersion because our sample distribution is non-Gaussian, and MAD is less sensitive to extreme values. Under this assumed model between birth radius and the \feh-\alphafe\ plane, this is consistent with an inside-out formation of the Milky Way; the older stars are more concentrated in the inner Galaxy. The younger stars on the other hand show mean distributions at larger radii and with wider distributions across  Galactic radii. 

Interestingly, we do not see any obvious age-\Rbirth\ trends when examining the data across all [Fe/H], i.e. without looking at different metallicity bins. This signal is erased, as the mean age distribution is a function of [Fe/H], so this age gradient, which is consistent with the idea of `inside-out' formation, is only seen when looking at the distribution of stellar ages in small ranges of [Fe/H] in our sample. In the pre-binned data (shown in Figure 3, panel (b), there is a clear density peak in the distribution in the \feh-\alphafe\ plane; this non-uniform density would presumably enable signatures in age and radius, which are correlated with this plane (i.e. \citealt{Hayden2017, Haywood2019}), without metallicity binning, as the majority of stars are at one particular metallicity already. The overall age gradient seen when examining all stars in the Milky Way (e.g. \citealt{Nessetal2016, Hayden2020}) is similarly presumably sensitive to the underlying density distribution of stars as a function of metallicity. This is an example of the  Yule-Simpson paradox, a phenomenon in which a trend appears in several groups of data, but disappears or reverses when the groups of data are combined. Examples of Yule-Simpson's paradox in Galactic archaeology can be found in \citet{Minchev2019YSP}. Additionally, samples with different metallicity are dominated by stars of different ages. As shown in \citet{Carrillo2022} Figure 2, the distribution of current radius R at low metallicity ([Fe/H]~=~-0.75) is dominated by 7-10~Gyr old stars, while the 1-3~Gyr old star population becomes the majority at high metallicity ([Fe/H]~=~0). This change in age dominance with [Fe/H] also appears in Figure \ref{fig:MetDist}. For the high metallicity sample, we are able to make a bin for 4-6 Gyrs stars but not for 10-12 Gyrs due to having too few old stars in the sample, and this is the opposite in the low metallicity sample. Therefore, we have to make bins according to [Fe/H] to account for the differing dominant age populations. In addition, this allows us to see inside-out growth in the level of chemical enrichment for mono-age populations. Comparing the distributions of the two 6-8 Gyr age bins (colored light pink) in both high and low metallicity samples, we find that the low metallicity sample peaks at a larger \Rbirth. A similar trend also exists in the distributions for age~=~8-10~Gyr stars (colored red). By selecting narrow metallicity bins, we show that the inside-out formation holds for different metallicities.

We summarise the birth radius-age relation, as shown in the top-left panel of Figure \ref{fig:KDEsummaryAge}. Overall, as the birth radius increases, the stellar age decreases. Similarly, as birth radius increases, the mean metallicity, [Fe/H], decreases, as shown in the bottom-left panel of Figure \ref{fig:KDEsummaryAge}. The top-right panel of Figure \ref{fig:KDEsummaryAge} shows the age dispersion as a function of birth radius. We see that small birth radii correspond to the highest age dispersions. Similarly, in the bottom-right panel of Figure \ref{fig:KDEsummaryAge}, we see that the [Fe/H] dispersion is highest at the smallest radii.

\section{individual abundance distributions at different Birth Radii}
\label{eledisdis}

\begin{figure*}
\centering
\includegraphics[width=0.7\textwidth]{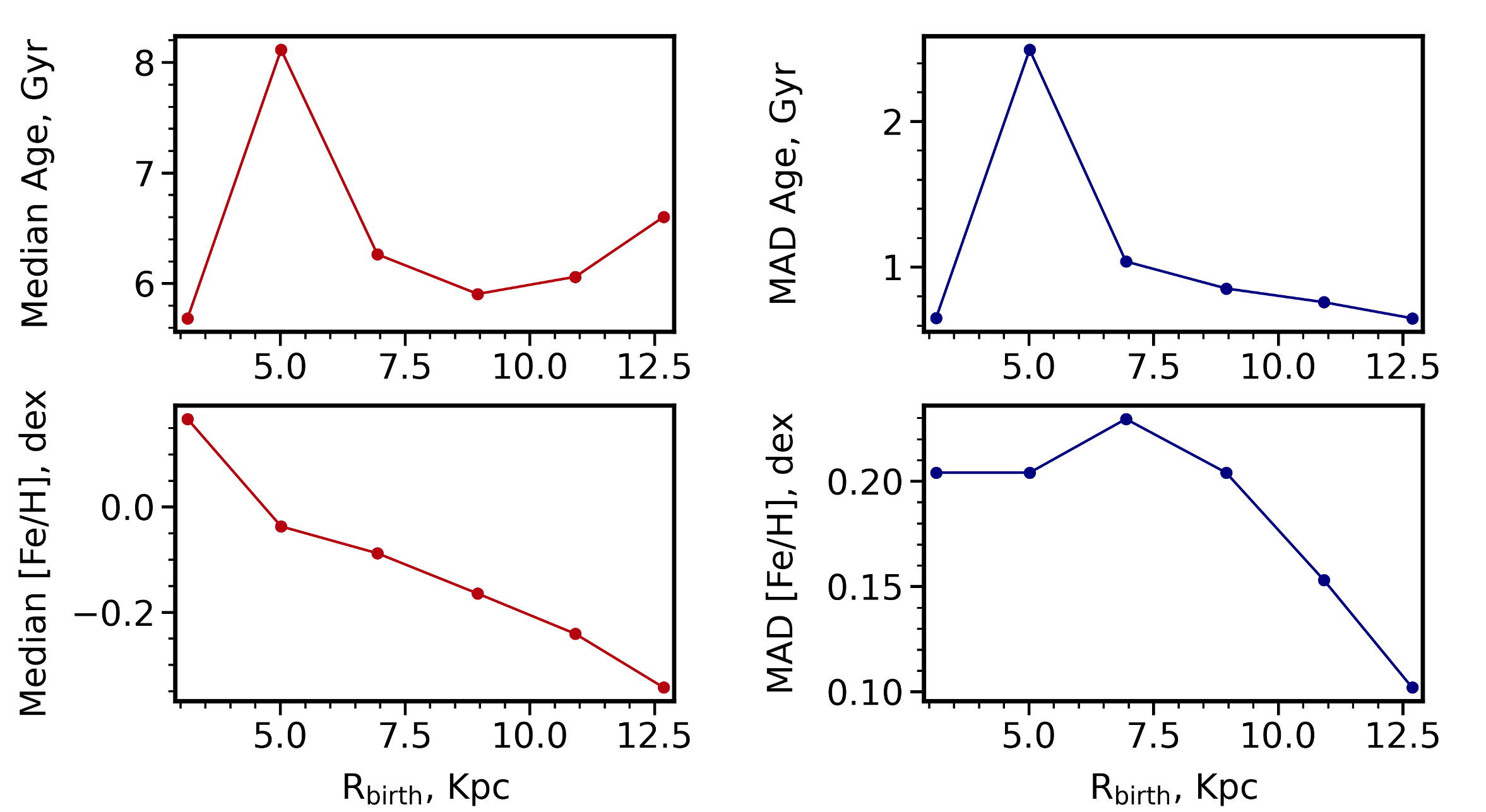} 
\caption{The median and median absolute deviation MAD of age (top row) and metallicity [Fe/H] (bottom row) vs. birth radius. Both decrease with increasing \Rbirth.} 
\label{fig:KDEsummaryAge}
\end{figure*}

\begin{figure*}
\centering
\includegraphics[width=0.8\textwidth]{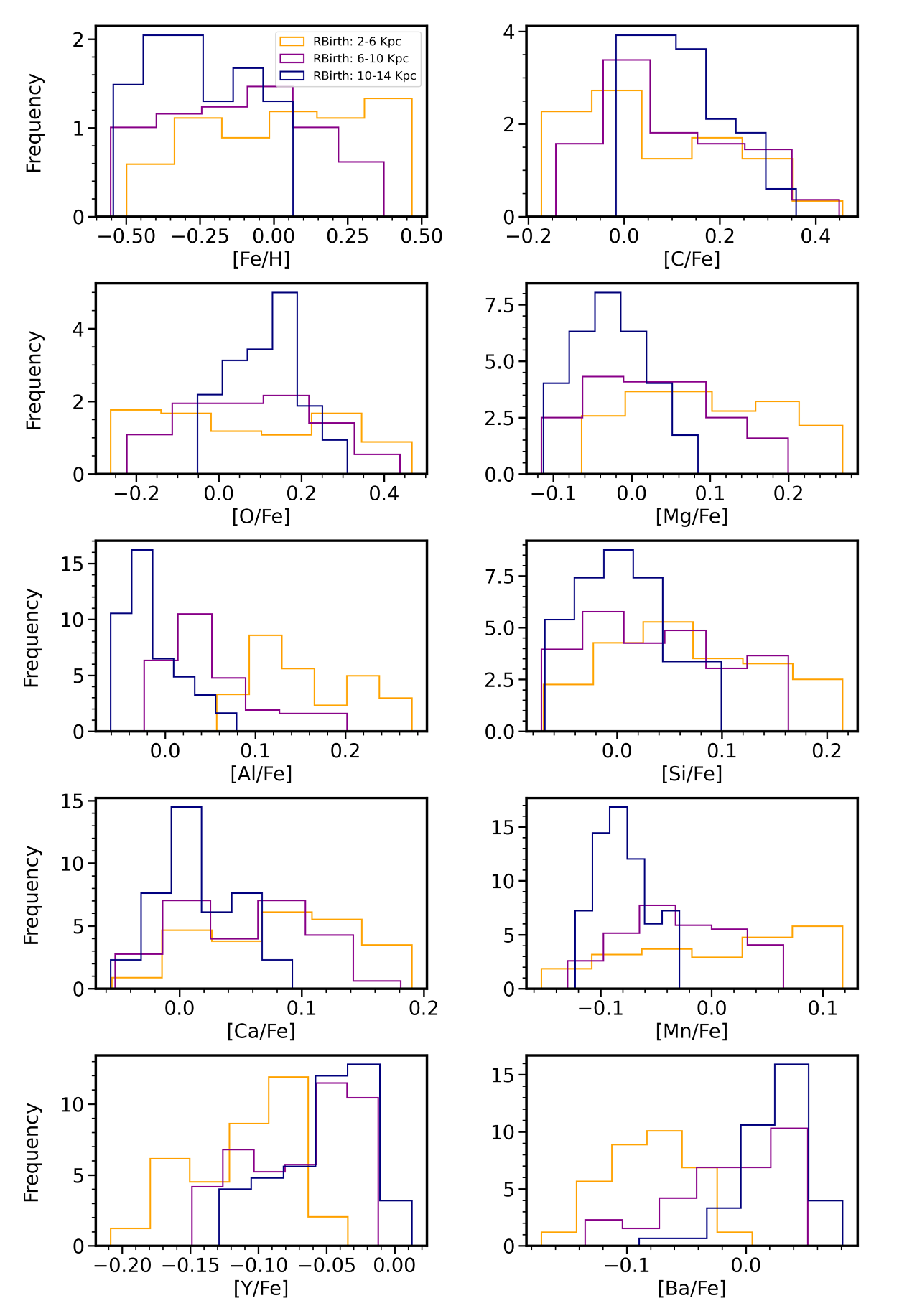} 
\caption{[X/Fe] histogram for data separated in 3 different birth radius bins, with bin width = 4 kpc. Small \Rbirth\ bins are colored in the lightest hue, and large \Rbirth\ bins are colored in the darkest hue. In all 10 panels, the elements exhibit distinct distribution trends at different \Rbirth\ bins, especially in the means and dispersion. Generally, we see that the distribution of smaller \Rbirth\ bins have larger dispersion, whereas larger \Rbirth\ bins show smaller dispersion. The median value of each element shows a gradient with \Rbirth. } 
\label{fig:XfeKDE}
\end{figure*}

We investigate the abundance distributions for the elements C, O, Mg, Al, Si, Ca, Mn, Fe, Y, and Ba, spanning the $\alpha$, odd-z, iron-peak, and neutron-capture groups of elements, at different birth radii. These [X/Fe] distributions are shown in Figure \ref{fig:XfeKDE}. The number of data points from Figure 4 (c) in each of the birth radius bins is 84 (2-6 kpc), 84 (6-10 kpc), and 53 (10-14 kpc).

We find a bimodal distribution towards small birth radius bins. High precision observational measurements of \feh-\alphafe\ in the solar neighborhood show a bimodality termed the `low' and `high' alpha discs (e.g. \citealt{Bensby2003,Reddy2003}). Across a wider Galactic radius range these change in their density contribution; the high-alpha sequence is concentrated to the inner Galaxy and the low-alpha sequence extends to the outer Galaxy (e.g. \citealt{Hayden2015}). The sampling we use for our analysis is evenly spaced across the full \feh-\alphafe\ plane as shown in Figure \ref{fig:GalahDist} (c). However, when we examine the individual abundance distributions, a bimodality appears in a number of individual elements at the smallest birth radii. This is presumably due to the contribution from both the high and low alpha discs at fixed birth radius in the inner Galaxy. In effect, this is a strong prediction of our model, that the disc is bimodal in elements at small birth radius. Furthermore, most of the elements show that the [X/Fe] distribution changes from wide (2-6 kpc) to narrow (10-14 kpc) as the birth radius increases.

\textbf{Metallicity, [Fe/H]}: The metallicity distribution at a small birth radius has a higher mean value. A decreasing mean metallicity gradient is observed with present-day guiding radius from the Milky Way center to the outer region (e.g. \citealt{Eilers2022}). This is inherited from a birth gradient in the gas metallicity (e.g. \citealt{Minchev2018, Chiappini2001}) but has presumably been weakened by radial migration over time (e.g. \citealt{Eilers2022}).

\textbf{Carbon}: Carbon is mainly produced in massive stars, followed by low-mass AGB stars \citep{Kobayashi2020}. Therefore, carbon distributions should be similar to that of $\alpha$-elements, as the majority of the $\alpha$-elements are produced in massive stars. The age-abundance relation for carbon in other observational works (e.g.\citealt{BensbyFeltzing2006, Ness2019}) shows a positive gradient, indicating that [C/Fe] is larger for older stars. In this study, the carbon abundance [C/Fe] has little relation with \Rbirth\ value. We see a weak and opposite trend where there is a slight shift in peak position (i.e. larger \Rbirth\ bins have greater peak [C/Fe]). However, Carbon changes over the evolution of the star due to dredge-up, so perhaps this is representative of the impact of the intrinsic evolution of the element rather than extrinsic (ISM).

\textbf{Oxygen, Magnesium, Silicon, and Calcium ($\alpha$-elements)}: For the $\alpha$-elements Mg, Si, and Ca, the distributions peak at a smaller mean [X/Fe] as the birth radius increases. The $\alpha$-elements are mainly produced through Type II Supernovae and their relative ISM contribution is diluted by the increasing supernovae Ia iron-peak pollution. Therefore, we expect the abundance of $\alpha$-elements, as a function of iron, to be lower in younger stars. We note that the oxygen abundance [O/Fe] shows the smallest evolution across different birth radii. The distribution is wider at smaller birth radii, and each of the distributions overlaps significantly. We see little variation in [O/Fe] with \Rbirth, which contradicts the progression found in other works (e.g. \citealt{DelgadoMena2019, Kobayashi2020}).

\textbf{Manganese (iron-peak)}: The iron-peak element Mn has a higher mean [Mn/Fe] value toward smaller birth radii. The iron-peak elements like Mn are generally synthesized in Type Ia supernovae and also in collapse supernovae. At the center of the Milky Way, younger stars are formed from more enriched gas compared to the outskirts of the Galaxy. As [Mn/Fe] increases with [Fe/H] (e.g. \citealt{Kobayashi2020}), [Mn/Fe] is expected to be higher in the Galactic center compared to that in the outskirts. In the age-abundance trends of Mn examined by \citet{Bedell2018} and \citet{Lu2021}, we see that both studies reveal a relatively flat but still positive age-abundance slope. In general, our result agrees with those from the previous studies.

\textbf{Aluminum (Odd-z)}: The odd-z element Al also has a higher mean abundance at smaller birth radii. Based on the prediction from the chemical evolution model of the Milky Way, [Al/Fe] decreases with time for stars with age 12 Gyrs and younger (\citealt{Horta2021} Figure 2). Since the majority of our sample stars are younger than 12 Gyrs, we expect our sample to behave similarly (i.e. decreasing [Al/Fe] with time). Moreover, \citet{Ness2019} examined the age-abundance relation of stars at solar metallicity and discovered a positive relation wherein [Al/Fe] increases with increasing age. Such a trend is also seen by \citet{Bedell2018} in their analysis of the Sun-like stars in the solar neighborhood. Thus, [Al/Fe] is expected to increase with decreasing birth radii, as predicted by both the chemical evolution model and the age-[Al/Fe] relation, and as shown in our results. Interestingly, the 2-6 birth radius bin does not follow the general trend of increasing dispersion with smaller birth radii. However, this is because, for all the binned data points with Al abundance available, the ones that fall in the 2-6 kpc birth radius bin do not span a wide range in [Fe/H], and thus the dispersion of the bin is smaller.

\textbf{Barium \& Yttrium (Neutron-capture)}: The two neutron-capture elements, Ba and Y, though centered on different values, have similar abundance distributions for stars at different birth radii; that is, the distribution peaks at a larger [X/Fe] value as birth radius increases. They exhibit an opposite trend as the aforementioned elements C, O, Mg, Al, Si, Ca, and Mn. This trend is consistent with the age-abundance relation for the neutron-capture elements from the literature (e.g. \citealt{Bedell2018}). According to the negative age-[Ba/Fe] relation \citep{DelgadoMena2019, Horta2022} as well as the age-birth radius relation (Figure \ref{fig:KDEsummaryAge} top-left panel), the older population were born at smaller mean birth radii with a lower [Ba/Fe] value. It is reassuring that Ba and Y have abundance distributions that behave similarly to birth radius, as both are considered s-process elements.

\begin{figure*}
    \includegraphics[width=0.8\textwidth]{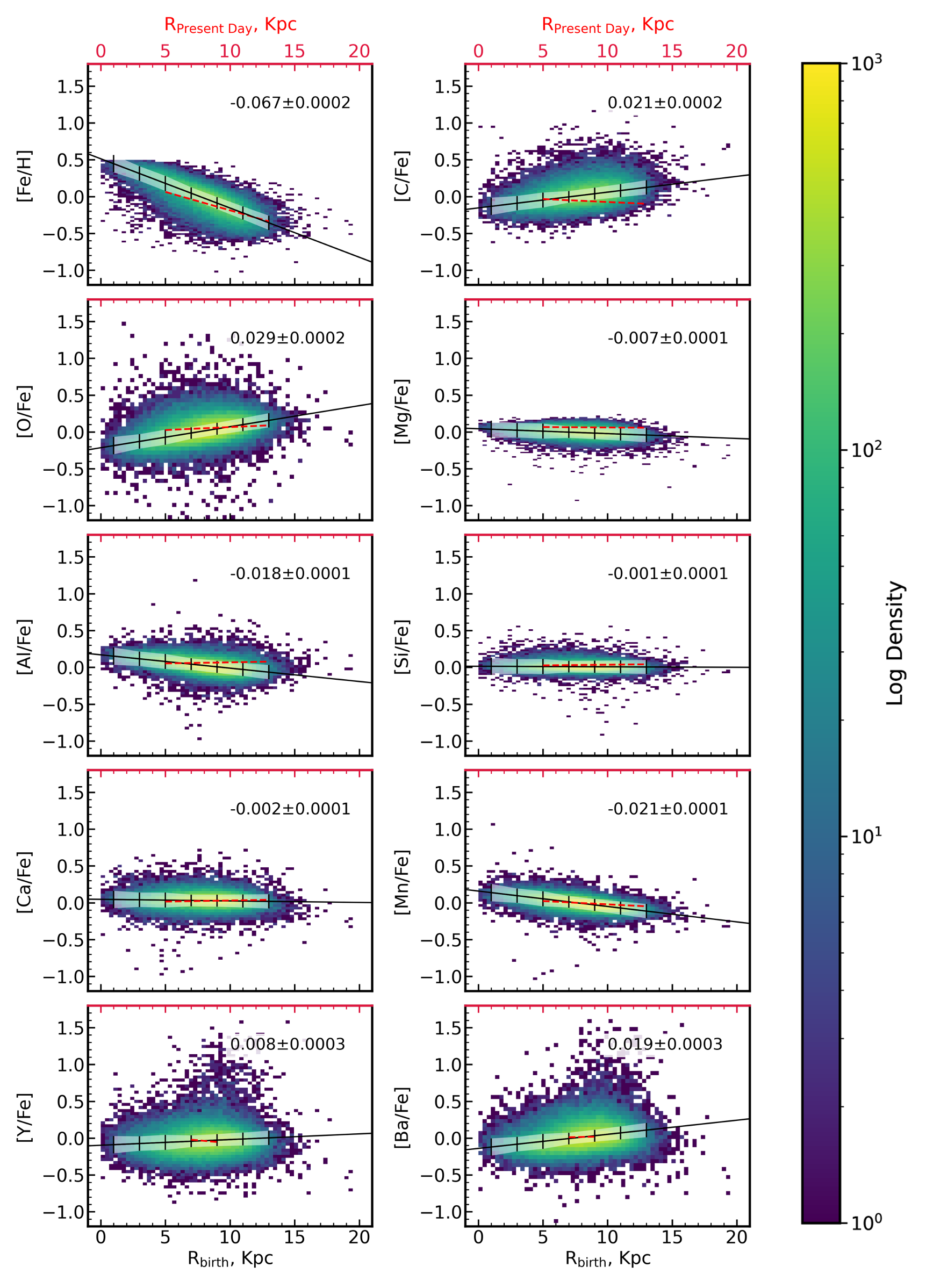} 
    \caption{Element abundance [X/Fe] vs. birth radius \Rbirth\ for GALAH DR3 low-$\alpha$ stars, colored by log density. The best-fit [X/Fe]-\Rbirth~gradient is shown with the black line, and these gradients are summarized in Table \ref{tab:lowalpha} column 3. The vertical error bars represent the MAD of [X/Fe] for stars in different \Rbirth~bins, with a bin width of 2 kpc. The red-dashed lines reflect the present-day gradients obtained using APOGEE DR17 (as listed in Table \ref{tab:lowalpha} column 2). For Y and Ba which there are no abundance measurements in APOGEE DR17, the red-dashed lines are their GALAH DR3 present-day gradients as shown in column 4.}
    \label{fig:KDEsummary}
\end{figure*}

Furthermore, we calculate and tabulate the \Rbirth-[X/Fe] gradients for the low-$\alpha$ stars. In Figure \ref{fig:KDEsummary}, we present the [X/Fe] vs. \Rbirth~plots for the low-$\alpha$ in GALAH DR3, with the black lines representing the best-fit gradients and colored by log density. The vertical error bar reflects the MAD of [X/Fe] in small \Rbirth\ bins with bin width = 2 kpc. The gradient results are summarized in Table \ref{tab:lowalpha} column 3. The reason we focus on the low-$\alpha$ population is that they exhibit the strongest change in element abundances across radius, but for the high-$\alpha$ stars, there is no obvious abundance trend associated with radius (e.g. \citealt{Eilers2022}). Adopting \citet{Griffith2022} cuts for low-$\alpha$ stars ([Mg/Fe]$>$0.12-0.13[Fe/H] if [Fe/H]$<$0; [Mg/Fe]$>$0.12 if [Fe/H]$>$0), the number of the low-$\alpha$ stars in our sample is $\sim56,000$. The inner-most \Rbirth\ bin (i.e. \Rbirth\ $<$5 kpc) seems to be an outlier to the general trend (referring to Figure \ref{fig:KDEsummaryAge}). Therefore, to justify a linear fit and gradient metric, we excluded the inner-most \Rbirth\ data points in our gradient calculations. The gradients are calculated over a \Rbirth\ range of 5-13 kpc. The largest abundance gradient with \Rbirth\ is seen in [Fe/H] at $-0.067$ dex/kpc followed by the individual element [O/Fe] with an [X/Fe]-\Rbirth\ slope of 0.029$\pm$0002~dex/kpc.

We emphasize that in the GALAH sample we use, our present-day radius is limited to the solar neighborhood, with a mean present-day radius of $8.14\pm0.35$ kpc. However, as stars migrate from birth, this survey still gives us access to stars born all over the disc, as parameterized in our model of \Rbirth\ (from 2-14 kpc). In APOGEE, the survey spans a present-day Galactocentric radius of 0.01-20 kpc, so we can directly compare and contrast our results for birth radius to the present-day radius with APOGEE. For example, Table \ref{tab:lowalpha} column 1 shows the abundance gradients for APOGEE DR16 low-$\alpha$ disc stars (i.e. [$\alpha$/M]<0.12, |z|<1) with current radius in the range of 5-13 kpc, obtained from \citet{Eilers2022} Figure 7. 
We show the seven elements [X/Fe] where X=C, O, Mg, Al, Si, Ca, and Mn) in APOGEE DR17 \citep{Majewski2017, Abdurro2022} that are in common with the elements used in this study. We also calculate gradients for the element abundances [X/Fe] independently, using $\sim 63,000$ APOGEE DR17 low-$\alpha$ stars. We adopt similar cuts as \citet{Eilers2022} (i.e. 4800~K$<$~\teff~$<$5800~K, \logg $<3.6$, [$\alpha$/M]<0.12, and |z|$<1$). The APOGEE gradients are summarized in Table \ref{tab:lowalpha} columns 1 and 2. In column 4, since GALAH covers a narrow range in current radius compared to APOGEE, the present-day radius-abundance gradients for GALAH low-$\alpha$ stars around the solar neighborhood only ($7<R<9$ kpc) are shown. We discuss these gradient comparisons in more detail in Section \ref{Discussion} below.

\section{Discussion}
\label{Discussion}

\begin{table*}
	\centering
	\caption{Current and birth radius [X/Fe] gradients for elements, restricted to low-$\alpha$ stars with |z|<1. \citet{Eilers2022} current radius-[X/Fe] gradients were obtained from their Figure 7 (from their APOGEE DR16 calibrated abundances).}
	\label{tab:lowalpha}
	\begin{tabular}{c||cccc}
            \hline
            & \citet{Eilers2022} & APOGEE DR17 & GALAH DR3 & GALAH DR3\\
            Element & Current (5-13 kpc) & Current (5-13 kpc) & Birth (5-13 kpc) & Current (7-9 kpc)\\
            \hline\hline
            Fe & -0.058 & -0.050 $\pm$ 0.0005 & -0.067 $\pm$ 0.0002 & -0.029 $\pm$ 0.003\\
            C & -0.009 & -0.008 $\pm$ 0.0002 & 0.021 $\pm$ 0.0002 & -0.001 $\pm$ 0.002\\
            O & 0.002 & 0.008 $\pm$ 0.0002 & 0.029 $\pm$ 0.0002 & 0.006 $\pm$ 0.002\\
            Mg & 0.002 & -0.001 $\pm$ 0.0002 & -0.007 $\pm$ 0.0001 & 0.001 $\pm$ 0.001\\
            Al & -0.002 & 0.003 $\pm$ 0.0002 & -0.018 $\pm$ 0.0001 & -0.003 $\pm$ 0.001\\
            Si & 0.003 & 0.002 $\pm$ 0.0001 & -0.001 $\pm$ 0.0001 & -0.001 $\pm$ 0.001\\
            Ca & 0.004 & 0.003 $\pm$ 0.0001 & -0.002 $\pm$ 0.0001 & 0.002 $\pm$ 0.001\\
            Mn & -0.014 & -0.009 $\pm$ 0.0001 & -0.021 $\pm$ 0.0001 & -0.009 $\pm$ 0.001\\
            Y & ... & ... & 0.008 $\pm$ 0.0003 & -0.015 $\pm$ 0.002\\
            Ba & ... & ... & 0.019 $\pm$ 0.0003 & 0.004 $\pm$ 0.002\\
		\hline
	\end{tabular}
\end{table*}

In this work, we explore the element abundance distributions of stars as a function of birth radius which we inferred from the [Fe/H]-[$\alpha$/Fe] plane alone, as motivated by cosmological simulations. We now discuss the validity of our assigned \Rbirth~tracks and the implications of our \Rbirth~estimates on the star formation history of the Galaxy. 

We test two other models for assigning the birth radius. We lay down horizontal and vertical \Rbirth \ tracks, on the \alphafe \ vs. \feh \ plane. From these alternate \Rbirth \ tracks, we produce [X/Fe] distributions of these stars with different \Rbirth, similar to Figure~\ref{fig:XfeKDE}. In the horizontal \Rbirth \ assignment, we see that the mean [Mn/Fe] and [Fe/H] values increase with increasing \Rbirth. This contradicts the observed [Fe/H] gradient (i.e. higher [Fe/H] at the center) of the Galaxy due to inside-out formation and therefore its longer history of star formation. In addition, there is no obvious trend in the dispersion across different \Rbirth\ bins for C, O, Al, Mn, Y, and Ba. As for the vertical \Rbirth \ assignment, the mean abundances of all four $\alpha$-elements, O, Mg, Si, and Ca, increase with \Rbirth, which does not agree with what is observed with the present-day guiding radius. Observations show that as radius increases the low-$\alpha$ populations dominate and in the inner Galaxy the high-$\alpha$ population has the highest density  (e.g. \citealt{Hayden2015, Lu2021}). Therefore, the alternative models we propose result in [X/Fe] distributions that are inconsistent with that of observations of present-day guiding radius. However, in general the  \Rbirth \ assignments motivated by the NIHAO-UHD simulations give rise to trends in the individual abundances [X/Fe] that are consistent with observations of element abundance distributions with present-day guiding radius. We have the expectation that the element abundance gradients and dispersions as a function of birth radius will be higher amplitude than that of the present-day guiding radius due to the impact of radial migration. Therefore, this gives us a better insight into the element abundance distributions at stellar birth place and time in the Milky Way disc.

Due to radial migration \citep[e.g.][]{Frankel2018}, we expect gradients in [X/Fe]-\Rbirth\ to be weakened over time. Therefore, abundance gradients across \Rbirth\ should be steeper than present-day gradients. This is indeed what we find for most elements.

Using the APOGEE DR16 data, \citet{Eilers2022} report negative present-day gradients across radius in the low-$\alpha$ disc (i.e. [$\alpha$/M]<0.12, |z|<1) for [Fe/H], as well as as the individual elements [X/Fe] where X = C, Al, Mn. For the elements X = O, Mg, Si, Ca they report positive gradients with Galactic radius. These gradients are summarised in Table \ref{tab:lowalpha} column 1. In column 2 of this table, we report the present-day abundance gradients we calculate with APOGEE. We find good agreement with the \citet{Eilers2022} analysis with the exception of a few elements. We note that the [Mg/Fe] and [Al/Fe] present-day abundance gradients are opposite in sign compared to \citet{Eilers2022} gradients. However,  the gradients for these two elements are very shallow. Some differences are not unexpected as we use the ASPCAP abundances from APOGEE and the \citet{Eilers2022} paper uses a data-driven approach to report calibrated abundances that these gradients are based on. Similarly, we report the present-day gradients in GALAH (column 4) for the low-alpha stars (adopting \citealt{Griffith2022} cuts). Note that the GALAH present-day gradients are over a restricted radius range, compared to APOGEE. Again there are some differences, and the GALAH gradients are shallower than APOGEE gradients.

The present-day element abundance gradients with radius in columns 1, 2, and 4 of Table 1 serve as a comparison to our calculated birth radius gradients (column 3).  We find that the GALAH birth radius gradients are steeper than both the GALAH present-day local gradient (column 4) and APOGEE present-day gradients (with wider present-day radius range; column 1 \& 2). The magnitude of the change in gradients varies with elements. 

We can therefore infer from our comparisons between columns 1 and 3 that gradients between elements and radius flatten over time. The element [Fe/H] shows the steepest gradient of $-0.067$ dex/kpc across birth radius. This flattens the order of 13 percent, to $-0.058$ dex/kpc from birth to present-day radius, well in agreement with recent theoretical predictions \citep{Buck2023}. The elements [X/Fe] where X = O, C, Mn, and Al all have the next steepest gradients from $-0.21$ dex/kpc to $0.029$ dex/kpc with birth radius. These flatten by between $\approx$ $0.02-0.03$ dex/kpc such that the present-day gradients for these elements vary between $\approx$ $-0.014-0.002$ dex/kpc. We also note that some of the gradients change sign, between birth and present-day radius (i.e. C, Mg, Ca, and Y). Similar [X/H] radial gradients being flattened over time is also observed in \citet{Ratcliffe2023}, in which they used an empirical approach from \citet{Lu2023} to derive \Rbirth~estimates for APOGEE DR17 red giant stars based on their age and [Fe/H].

The individual abundances of stars as a function of birth radius record the star-forming environment at that location and time in the disc.  A recent study by \citet{Horta2022} employed chemical evolution modeling \citep{Rybizki2017} to use ages and individual abundances of GALAH stars to infer environmental parameters (i.e. high-mass slope of the \citealt{Chabrier2001} IMF ($\alpha_\text{IMF}$), number of SN~Ia exploding per solar mass over 15 Gyr ($\log_{10}(\text{SNIa})$)). Their analysis assumed a link between using small bins in [Fe/H]-[Mg/Fe]-[Ba/Fe]-age for the chemical evolution model, as representative of linking to the interstellar medium conditions at different birth radii. They subsequently examined the model parameter gradients across present-day radius. They found that the abundances give rise to a gradient in the high-mass end of the disc's initial mass function. They report that this is more top-heavy towards the inner disc, and more bottom-heavy in the outer disc. Using our birth radius assignment, it would be possible to directly infer the environmental parameters as a function of birth radius and compare the conditions at different birth places and times in the star-forming disc directly.

\section{Conclusion}
\label{sec:conclusion}
This work examines the distribution of individual abundances [X/Fe] of elements C, O, Mg, Al, Si, Ca, Mn, Y, and Ba for disc stars in different birth radii. To do this, we assumed seven birth radius tracks across the \alphafe \ vs. \feh \ plane of $\sim~59,000$ GALAH DR3 disc stars and assigned each star a birth radius. This formalism is based on the  NIHAO-UHD simulations \citep{Buck2020} (see Figures \ref{fig:Buck2020} and \ref{fig:GalahDist}). We emphasize that our adopted model of birth radius is not calibrated to quantitatively map a location in the \feh-\alphafe\ plane to the birth radius. Rather, this serves as a tool to trace the element abundance and age distribution of stars across the disc from their origin. Via this approach, we can map variations in time of birth and in individual channels of enrichment to differences in the star-forming environment over time and radius. Below we summarize our main results:

\begin{itemize}
    \item {The \Rbirth~distribution as a function of age supports an inside-out growth for the Milky Way disc (Figure \ref{fig:MetDist}). There is a larger mean value in \Rbirth~for the younger population (i.e., $\sim$10 kpc) compared to the older population (i.e., $\sim$4 kpc). This result is consistent with a number of earlier studies (e.g. \citealt{Bensby2011, Bovy2012, Minchev2018}).}
    \item {The \Rbirth~distribution dispersions change with age as well i.e., the median absolute deviation changes from 0.8 kpc to 1.5 kpc going from older to younger stellar populations as the Milky Way disc grows with time and therefore has star formation over a larger region.}    
    \item{There is a clear progression in the median [X/Fe] trend with \Rbirth: Mg, Si, Ca, Mn, and Al all decrease while C, O, Y, and Ba all increase with increasing \Rbirth.}
    \item{For the low-$\alpha$ population, the abundance gradients are steeper in birth compared to present-day radius. The [Fe/H]-\Rbirth\ gradient measures $-0.067\pm0.0002$ dex/kpc compared to the [Fe/H] present-day gradient of -0.058 dex/kpc. The [O/Fe] abundance is the next strongest indicator of \Rbirth; it exhibits the steepest [X/Fe]-\Rbirth\ slope of the [X/Fe] measurements (see Table \ref{tab:lowalpha}) and is $0.029\pm0.0002$~dex/kpc in \Rbirth~and 0.002 dex/kpc in present-day.}
\end{itemize}

We tested two other birth radius assignments based on stars' location on the \alphafe\ vs. \feh \ plane, but neither return physically plausible [X/Fe] distributions across radius. Furthermore, our model adopted from the simulation gives sensible results that are aligned with expectations. For example, according to radial migration, we expect birth radius gradients to be steeper in the past, which we find. Therefore, the adopted model for birth radius appears physically plausible and presumably gives insight into the relative distribution of individual abundances across the disc as it formed. Our model uses no direct information about present-day birth radius and is also therefore a useful comparison to models that do assume a relationship with the present-day radius.

In summary, aided by \Rbirth~tracks inspired from a cosmological hydrodynamical simulation of a Milky Way-like galaxy and assumptions constrained to the \alphafe\ vs. \feh \ plane, we are able to recover the inside-out growth of the Milky Way disc and the \textit{spatial} evolution in its chemical abundance distributions. This work serves as a proof of concept of the legitimacy of this modeling approach, and in future it can be applied to additional large spectroscopic survey data. This includes data that covers a larger area of the disc, such as SDSS-V Milky Way Mapper. In addition, chemical evolution modeling would add another dimension in investigating the validity of these \Rbirth~assignments (e.g. \citealt{Buck2021}). Nonetheless, this work shows that assigning birth radius to stars in the Milky Way and studying the element abundance distributions over time and birth place is very promising. This is demonstrative of the utility in using ensembles individual abundances to trace the formation of the Milky Way disc.

\section*{Acknowledgements}
AC acknowledges support from the Science and
Technology Facilities Council (STFC) [grant number
ST/T000244/1] and the Leverhulme Trust. 
TB's contribution to this project was made possible by funding from the Carl Zeiss Stiftung.

\section{Data Availability}
The GALAH DR3 data used in this article are available at \href{https://www.galah-survey.org/dr3/the_catalogues}{https://www.galah-survey.org/dr3/the\_catalogues}. The APOGEE DR17 data used in this article are available at \href{https://www.sdss4.org/dr17}{https://www.sdss4.org/dr17}. Simulation data from the NIHAO-UHD project is availabel at \href{https://tobias-buck.de/#sim_data}{https://tobias-buck.de/\#sim\_data}. Other data used in this article can be made available upon reasonable request to the corresponding authors.

\bibliographystyle{mnras}
\bibliography{example}

\appendix
\section{Element Abundance vs. Effective Temperature}
\begin{figure}
\centering
\includegraphics[width=0.48\textwidth]{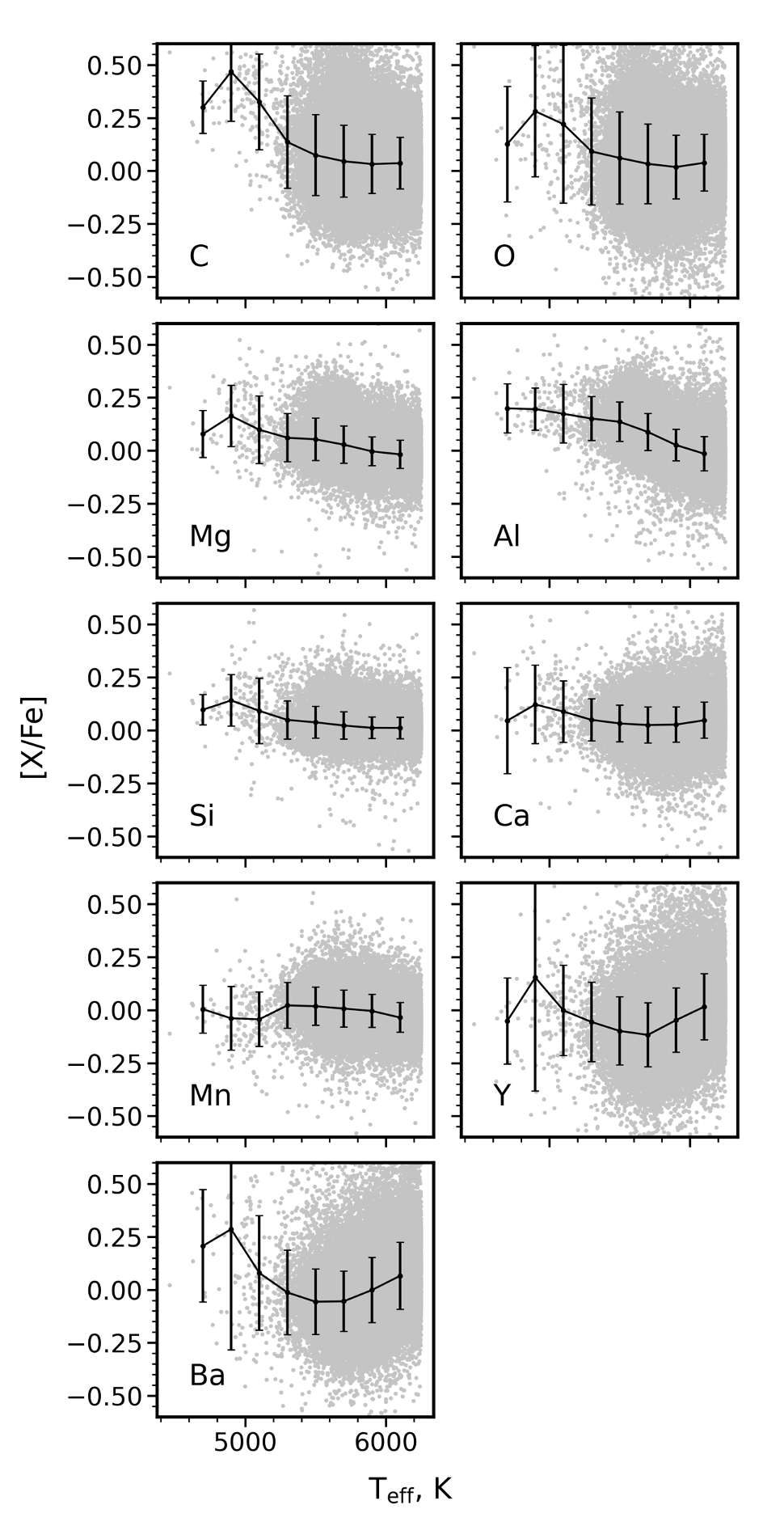} 
\caption{Element abundance [X/Fe] vs. effective temperature \teff. The mean and scatter in shown in black solid dots as a function of \teff. We see no obvious [X/Fe]-\teff \ trends.}
\label{fig:RunnningMean}
\end{figure}

We include here the element abundance [X/Fe] vs. effective temperature \teff \ plot to show that there is no trend associated with [X/Fe] and \teff, see Figure \ref{fig:RunnningMean}.

\bsp
\label{lastpage}
\end{document}